\documentclass[%
prl,
superscriptaddress,
 amsmath,amssymb, nofootinbib,
 aps,
]{revtex4-2}

\pdfoutput=1

\usepackage[utf8]{inputenc}

\usepackage{xcolor}

\usepackage{graphicx}
\usepackage{dcolumn}
\usepackage{bm}
\usepackage{hyperref}

\newcommand{\rf}[1]{Eq.~(\ref{#1})}

\newcommand{\nn}{\nonumber}
\newcommand{\mysection}[1]{{\vspace{10 pt}\noindent \emph{{\textbf{#1}}--}}}

\newcommand{\supplemental}[1]{{\vspace{10 pt}\centerline{\textbf{#1}}}}

\newcommand{\be}{\begin{eqnarray}}
\newcommand{\ee}{\end{eqnarray}}

\newcommand{\p}{\partial}

\newcommand{\D}{{\mathcal D}}
\newcommand{\A}{{\mathcal A}}
\newcommand{\Q}{{\mathcal Q}}
\newcommand{\N}{{\mathcal N}}
\newcommand{\q}{\mathcal{J}}
\newcommand{\f}[2]{\frac{#1}{#2}}

\newcommand{\half}{{\frac{1}{2}}}

\newcommand{\vv}{{\bf v}}
\newcommand{\kk}{{\bf k}}
\newcommand{\ttt}{\tau_R}

\newcommand{\xa}{\Xi_1}

\newcommand{\xb}{\Xi_2}

\allowdisplaybreaks

\begin{document}

\author{Jorge Noronha}
\email{jn0508@illinois.edu}
\affiliation{Illinois Center for Advanced Studies of the Universe\\ Department of Physics, 
University of Illinois at Urbana-Champaign, Urbana, IL 61801, USA}

\author{Micha\l\ Spali\'nski}
\email{michal.spalinski@ncbj.gov.pl}
\affiliation{National Centre for Nuclear Research, 02-093 Warsaw, Poland}
\affiliation{Physics Department, University of Bia{\l}ystok,
  15-245 Bia\l ystok, Poland}

\author{Enrico Speranza}
\email{espera@illinois.edu}
\affiliation{Illinois Center for Advanced Studies of the Universe\\ Department of Physics, 
University of Illinois at Urbana-Champaign, Urbana, IL 61801, USA}

\title{Transient Relativistic Fluid Dynamics in a General Hydrodynamic Frame}


\begin{abstract} 
  
We propose a new theory of second-order viscous relativistic hydrodynamics which
does not impose any frame conditions on the choice of the hydrodynamic
variables. It differs from Mueller-Israel-Stewart theory by including additional transient 
degrees of freedom, and its 
first-order truncation reduces to Bemfica-Disconzi-Noronha-Kovtun theory. Conditions for causality and
stability are explicitly given in the conformal regime.   As an illustrative example, 
we consider Bjorken flow solutions to our equations and identify variables which
make a hydrodynamic attractor manifest.


\end{abstract}

\maketitle

\mysection{Introduction} Nonrelativistic viscous hydrodynamics -- the
Navier-Stokes theory -- provides evolution equations for hydrodynamic variables with a vast range of
applicability. While the physics and mathematics of this theory remain challenging still today,
its formulation is textbook material \cite{LandauLifshitzFluids}. In contrast, the formulation of a suitable relativistic
generalization of Navier-Stokes theory remains an important topic of
current research. Strong motivation for these considerations comes from the quark-gluon plasma created in heavy-ion experiments \cite{Heinz:2013th}, which is described as a relativistic viscous fluid \cite{Romatschke:2017ejr}, and neutron star mergers, 
modeled using fluid dynamics in general relativity \cite{Baiotti:2016qnr}.

The most widely used  relativistic viscous hydrodynamic models are based on the approach of M\"uller, Israel, and Stewart (MIS)~\cite{MIS-1,MIS-6}.
These models overcome the difficulties with causality and stability found in relativistic Navier-Stokes theory \cite{Hiscock_Lindblom_instability_1985} by incorporating transient, non-hydrodynamic degrees of freedom in addition to long-lived hydrodynamic modes. In a recent development, an alternative 
was discovered by Bemfica, Disconzi, Noronha \cite{Bemfica:2017wps,Bemfica:2019knx,Bemfica:2020zjp} and Kovtun \cite{Kovtun:2019hdm,Hoult:2020eho} (BDNK), who introduced a nonhydrodynamic
sector 
in a different way than MIS. The basic observation of \cite{Bemfica:2017wps} and \cite{Kovtun:2019hdm} was that, since the hydrodynamic variables (such as temperature, flow velocity, and chemical potential) do not have a unique definition out of equilibrium \cite{MIS-6,Kovtun:2012rj}, one may take advantage of this freedom to formulate first-order hydrodynamics 
without imposing either the Landau \cite{LandauLifshitzFluids} or Eckart hydrodynamic frames \cite{EckartViscous}. Instead, one can define the theory in a general hydrodynamic frame without any additional fields beyond those found in ideal hydrodynamics and prove that it is causal and hyperbolic in the full nonlinear regime and stable at the linear level \cite{Bemfica:2017wps,Kovtun:2019hdm,Bemfica:2019knx,Hoult:2020eho,Bemfica:2020zjp}. 
The choice of hydrodynamic frame also affects causality and well-posedness of hydrodynamic theories in the presence of quantum anomalies \cite{Speranza:2021bxf}.

In this Letter we clarify the relationship between the MIS and BDNK approaches for the first time by showing that the MIS scenario can be consistently formulated without imposing any choice of frame, which leads to a hydrodynamic theory involving additional relaxation times. This theory contains more transient degrees of
freedom than either MIS of BDNK theories, and reduces to the latter when truncated to first order in gradients. We develop this new theory in the case where one has a conserved current along with the energy-momentum tensor, without making any special assumptions concerning the equation of state or the symmetries of solutions. We also discuss stability and causality of this theory as well as some special solutions in the case of conformal systems. 

\mysection{Hydrodynamics in a general frame} 
Let us consider the case of a nonconformal fluid described by an energy-momentum tensor $T^{\mu\nu}$ and a (e.g. baryon) current $J^\mu$. The most general decomposition of these tensors is 
\begin{align}
T^{\mu\nu} ={}& (\varepsilon+\mathcal{A})u^\mu u^\nu + (P + \Pi)\Delta^{\mu\nu} +\pi^{\mu\nu}  + \mathcal{Q}^\mu u^\nu + \mathcal{Q}^\nu u^\mu , \label{defineTmununc}\\
\label{defineJmu}
J^\mu ={}& (n+\N)u^\mu + \q^\mu ,
\end{align}
where $\varepsilon$ and $n$ are the equilibrium energy and number density, respectively, $P=P(\varepsilon,n)$ is the equilibrium pressure which defines the equation of state,  $u^\mu$ (with $u_\mu u^\mu = -1$) is the 4-velocity of the fluid, and $\Delta^{\mu\nu}=u^\mu u^\nu + g^{\mu\nu}$. The quantities $\pi^{\mu\nu}$, $\Pi$, $\Q^\mu$ and $\q^\mu$ are the shear-stress tensor, bulk-viscous pressure, energy flux vector and diffusion current, respectively, while $\mathcal{A}$ and $\N$ are the out-of-equilibrium corrections to the energy density and charge density, respectively. The resulting conservation equations $\partial_\mu T^{\mu\nu}=0$ and $\partial_\mu J^{\mu}=0$ can be written as
\begin{subequations}
\begin{align}
    {D}(\varepsilon+\mathcal{A})+ (\varepsilon + P +\mathcal{A}+\Pi)\theta+\pi_{\mu\nu}\sigma^{\mu\nu}+\partial_{\mu}\mathcal{Q}^\mu +\mathcal{Q}_\mu D u^\mu=0 , & \\
    (\varepsilon + P + \mathcal{A}+\Pi) Du^\alpha + \Delta^{\alpha\mu}\partial_{\mu} (P+\Pi) +\Delta^\alpha_\nu \partial_\mu \pi^{\mu\nu} + \mathcal{Q}^\mu \partial_\mu u^\alpha + \Delta^\alpha_\nu D \mathcal{Q}^\nu +\mathcal{Q}^\alpha \theta =0, &\\
    D(n+\N) +(n+\N)\theta + \partial_\mu \q^\mu =0, &
\end{align}
\label{eq:conservation}
\end{subequations}
with $D=u^\mu \partial_\mu$, $\theta =\partial_\mu u^\mu$, $\sigma^{\mu\nu} = \Delta^{\mu\nu\alpha\beta}\partial_\alpha u_\beta$, where  $\Delta^{\mu\nu\alpha\beta} = \left(\Delta^{\mu\alpha}\Delta^{\nu\beta}+\Delta^{\mu\beta}\Delta^{\nu\alpha}\right)/2-\Delta^{\mu\nu}\Delta^{\alpha\beta}/3$. The conservation equations need to be supplemented by evolution equations for the fields $\A$,
$\Pi$, $\Q^\mu$, $\pi^{\mu\nu}$, $\N$ and $\q^\mu$. Such equations can be derived in various ways assuming specific microscopic models, e.g. applying the method of moments in the relativistic Boltzmann equation \cite{Denicol:2012cn}.
We do not wish to restrict our arguments to any specific microscopic picture, so we adopt an effective theory approach where we write down an entropy current containing all terms allowed by symmetries up to second order in deviations from equilibrium. This makes it possible to apply the Israel-Stewart argument \cite{MIS-6} to derive evolution equations which guarantee that the second law of thermodynamics is locally obeyed, now in a general hydrodynamic frame. 

Taking into account the decomposition in Eqs.~\eqref{defineTmununc} and \eqref{defineJmu}, one is led to the following nonequilibrium entropy current
\begin{align}
    S^{\nu}_\text{neq} = &\left( s + \frac{\A}{T} -\frac\mu T \N\right) u^\nu + \frac{\mathcal{Q}^\nu}{T} - \frac \mu T \q^\nu  
    -\frac1 {2T}\left(\beta_\pi \pi^{\lambda\rho}\pi_{\rho\sigma}+ \beta_\Q \mathcal{Q}^\lambda \mathcal{Q}_\lambda +\beta_\A \mathcal{A}^2 +\beta_\Pi \Pi^2 +\beta_\N \N^2 +\beta_\q \q^\lambda \q_\lambda  \right. \nonumber\\
    & \left. - 2 \alpha_{\N\Pi} \N \Pi - 2 \alpha_{\N\A} \N\A - 2\alpha_{\Pi\A} \Pi \A - 2 \alpha_{\Q \q} \Q^\lambda \q_\lambda  \right)u^\nu ,
    \label{eq:entrcur}
\end{align}
where $T$ and $\mu$ are the locally-defined temperature and chemical potential, $s=(\varepsilon+P-\mu n)/T$ is the equilibrium entropy density, while $\beta$'s and $\alpha$'s are independent coefficients. We note that Eq.\ \eqref{eq:entrcur} is not the most general expression for the entropy current, as other nonequilibrium terms perpendicular to $u^\nu$ are also possible. For the general expression of the entropy current, and the resulting equations of motion, see the Supplemental Material. 
Entropy production is guaranteed to be non-negative when the following relaxation equations hold
\begin{subequations}
\label{eqs:relax_non}
\begin{align}
 \frac{\mathcal{A}}{\varphi T} = &-\frac{\beta_\A}{T} D \mathcal{A} + D\left(\frac1T \right) -\frac12 \mathcal{A} \partial_\nu \left( \frac{\beta_\mathcal{A}}{T}u^\nu \right)  
  + \frac{\alpha_{\mathcal{N}\mathcal{A}}}{T} D \mathcal{N} 
  + \frac{\alpha_{\Pi \mathcal{A}}}{T} D \Pi ,
  \\
 \frac{\Pi}{\zeta T} = &-\frac{\beta_\Pi}{T} D \Pi -\frac 1T \theta -\frac12 \Pi \partial_\nu \left( \frac{\beta_\Pi}{T}u^\nu \right) + \frac{\alpha_{\N\Pi}}{T} D \N  + \frac{\alpha_{\Pi \A}}{T} D \mathcal{A}  ,  \\
 \frac{\Q^\nu}{\psi T} = &-\frac{\beta_\Q}{T} \Delta^{\nu}_\lambda D \mathcal{Q}^\lambda + \Delta^{\nu}_\lambda \partial^{ \lambda}\left(\frac1T \right) - \frac1T Du^\mu -\frac12 \mathcal{Q}^\nu \partial_\lambda \left( \frac{\beta_\Q}{T}u^\lambda \right) + \frac{\alpha_{\Q \q}}{T} \Delta^\nu_\lambda D \q^\lambda ,
  \\
 \frac{\pi^{\nu\lambda}}{2\eta  T} = &-\frac{\beta_\pi}{T} \Delta^{\nu\lambda\delta\rho} D \pi_{\delta\rho} -\frac{\sigma^{\nu\lambda}}{T} -\frac12 \pi^{\nu\lambda} \partial_\rho \left( \frac{\beta_\pi}{T}u^\rho \right) , \\
 \frac{\N}{\xi T} = & -\frac{\beta_\N}{T} D \N - D \left(\frac\mu T \right) -\frac12 \N \partial_\nu \left( \frac{\beta_\N}{T}u^\nu \right)  + \frac{\alpha_{\N\Pi}}{T} D \Pi +  \frac{\alpha_{\N\A}}{T} D \A+  \Pi \partial_\nu \left( \frac{\alpha_{\N\Pi}}{T} u^\nu\right) 
   +  \mathcal{A} \partial_\nu \left( \frac{\alpha_{\N\A}}{T} u^\nu \right) ,  \\
 \frac{\q^\nu}{\kappa T}= & -\frac{\beta_\q}{T} \Delta^{\nu}_\lambda D \q^\lambda - \Delta^{\nu}_\lambda \partial^{ \lambda}\left(\frac \mu T \right)  -\frac12 \q^\nu \partial_\lambda \left( \frac{\beta_\q}{T}u^\lambda \right) + \frac{\alpha_{Q \q}}{T} \Delta^{\nu}_\lambda  D \mathcal{Q}^\lambda +  \mathcal{Q}^\nu \partial_\lambda \left( \frac{\alpha_{\Q \q}}{T} u^\lambda \right) ,
 \label{eq:relax2}
 \end{align}
\end{subequations}
where $\zeta$, $\eta$ are the bulk and shear viscosities, and $\kappa$ is the diffusion coefficient, while $\varphi$, $\psi$ and $\xi$ are new transport coefficients. The divergence of the entropy current is determined using the conservation laws, together with the relaxation equations above, which leads to 
\begin{align}
\partial_\mu S_\text{neq}^\mu = {}&\frac{1}{T}\left( \frac{\A^2}{\varphi} + \frac{\Pi^2}{\zeta} + \frac{\Q^\mu \Q_\mu}{\psi} +
\frac{\pi^{\mu\nu}\pi_{\mu\nu}}{2 \eta} 
 +
\frac{\N^2}{\xi} + \frac{\q^\mu \q_\mu}{\kappa} 
\right)  ,
\end{align}
so one can guarantee that it is always nonnegative by requiring that $\eta, \, \zeta, \, \kappa,\, \varphi,\, \psi,\, \xi$ are positive. Equations~\eqref{eqs:relax_non} provide a generalization of MIS in a general hydrodynamic frame. A kinetic theory derivation of similar equations was provided in \cite{Rocha:2021lze}. 


In the limit of vanishing $\varphi, \, \psi, \, \xi$ as well as the $\alpha$-parameters (with the remaining parameters finite), our new theory reduces to MIS in the Landau frame where $\A=\N=0$ and $\Q^\mu=0$. 
Note that even if one sets the initial values of $\A$, $\Q^\mu$ and $\N$ to zero, as in current heavy-ion simulations \cite{Romatschke:2017ejr}, in our theory those quantities will become nonzero as the system evolves. Therefore, in simulations of our theory
one could observe how the system dynamically deviates from the Landau frame.


\mysection{The gradient expansion}  For generic values of the transport coefficients, the derivative expansion of \rf{eqs:relax_non}, evaluated on solutions of the conservation equations
\eqref{eq:conservation}, leads to the following relations at first order in derivatives
\begin{widetext}
\begin{subequations}
\label{eqs:gradexp}
\begin{align}
\frac{\A}{\varphi}={}&\left(-\frac1 T + \beta_\A \frac{\partial \varepsilon}{\partial T} - \alpha_{\N\A} \frac{\partial n}{\partial T}  \right) D T  +\left[\beta_\A \frac{\partial \varepsilon}{\partial (\mu/T)} - \alpha_{\N\A} \frac{\partial n}{\partial (\mu/T)} \right] D \left( \frac\mu T \right) + \left[ \beta_\A (\varepsilon + P) - \alpha_{\N\A}(n + \N) \right] \theta , \\
\frac\Pi\zeta = {} & -  \left(\alpha_{\Pi A} \frac{\partial \varepsilon}{\partial T}+ \alpha_{\N\Pi} \frac{\partial n}{\partial T} \right) D T -  \left[\alpha_{\Pi A} \frac{\partial \varepsilon}{\partial (\mu/T)}+ \alpha_{\N\Pi} \frac{\partial n}{\partial (\mu/T)} \right] D \left( \frac\mu T \right) -  \left[ 1+ \alpha_{\Pi A}(\varepsilon +P) + \alpha_{\N\Pi} (n + \N)  \right]\theta , \\
\frac{\Q^\nu}{\psi} = {}&  \left(-\frac1 T + \beta_\Q \frac{\partial P}{\partial T}   \right) \Delta^\nu_\lambda\partial^{\lambda} T + \beta_\Q \frac{\partial P}{\partial (\mu/T)} \Delta^\nu_\lambda\partial^{\lambda} \left(\frac\mu T \right) + \left[ \beta_\Q (\varepsilon + P) -1 \right] D u^\nu , \\
\frac{\pi^{\nu\lambda}}{2\eta } ={}& - \sigma^{\nu\lambda} , \\
\frac{\N}{\xi} = {}& \left( -\alpha_{\N\A} \frac{\partial \varepsilon}{\partial T} + \beta_\N \frac{\partial n}{\partial T}  \right) DT + 
\left[ -\alpha_{\N\A} \frac{\partial \varepsilon}{\partial (\mu/T)} + \beta_\N \frac{\partial n}{\partial (\mu/T)} -T  \right]D \left(\frac \mu T\right) + \left[ -\alpha_{\N\A} (\varepsilon + P) + \beta_\N (n+\N)  \right]\theta, \\
\frac{\q^\nu}{\kappa} ={}& -\alpha_{\Q \q} \frac{\partial P}{\partial T} \Delta^\nu_\lambda\partial^{\lambda} T - \left[ T + \alpha_{\Q \q } \frac{\partial P}{\partial (\mu/T)}  \right] \Delta^\nu_\lambda\partial^{\lambda} \left(\frac \mu T \right) - \alpha_{\Q \q} (\varepsilon +P) D u^\nu , 
\end{align}
\end{subequations}
\end{widetext}
where we used $T$ and $\mu/T$ for the two independent thermodynamic variables. 
Therefore, one can see that the first-order truncation of our theory reduces to BDNK \cite{Bemfica:2017wps,Bemfica:2019knx,Bemfica:2020zjp,Kovtun:2019hdm,Hoult:2020eho}. 
Note that this requires including the mixed terms involving the $\alpha$-parameters in the entropy current, but additional terms are possible. However, we stress that the most general form of the equations of motion, given in the Supplemental Material, leads to the same expression for the gradient expansion truncated at first order in Eqs.\ \eqref{eqs:gradexp}. This is why we used the simpler expression for the nonequilibrium entropy current in \eqref{eq:entrcur}.

For applications of Eqs.\ \eqref{eqs:relax_non} in the description of relativistic flows, it is critically important to establish the domain of parameters for which these equations lead to causal and stable evolution, with a well-posed initial value problem. Causality conditions, valid in the nonlinear regime, can be obtained using the same tools employed in \cite{Bemfica:2020xym}, which investigated DNMR theory \cite{Denicol:2012cn} in the Landau frame at zero chemical potential. However, such analysis will be significantly more complex in the case under consideration, given the presence of finite chemical potential and additional viscous fluxes. Strong hyperbolicity, and consequently well-posedness of solutions, are much more challenging to establish and results are available only for simpler cases, such as the MIS/DNMR equations where only bulk viscosity is present \cite{Bemfica:2019cop}. Therefore, we leave a general analysis of causality and hyperbolicity for future work.

\mysection{The conformal case} We now focus on conformal fluids at zero chemical potential, so we set $J^\mu=0$. This special case is important for applications in ultrarelativistic heavy-ion physics. 
Conformal symmetry implies that the equation of state is $P =\varepsilon/3$ and  $\Pi=\mathcal{A}/3$. 
Furthermore, we define the effective
temperature $T$ by $\varepsilon \propto T^4$. The equilibrium entropy density is $s =
4\varepsilon/(3T)$.  
The conservation equations 
now read
\begin{subequations}
\label{eq:conservation0}
\be
\D(\varepsilon+\mathcal{A})+\pi_{\mu\nu}\sigma^{\mu\nu}+\mathcal{D}_{\mu}\mathcal{Q}^\mu
&=&0, \;\;\;\;\;\\
\label{conservation2}
 \Delta^{\lambda\nu} 
 \left[
 \f{1}{3}\D_{\lambda}(\varepsilon + \mathcal{A}) +
  \D^\mu\pi_{\mu\lambda}
\right] 
+  \mathcal{Q}^\mu \mathcal{D}_\mu u^\nu +
  \D\mathcal{Q}^{\nu} &=& 0, \;\;\;\;\;
\ee
\end{subequations}
where $\mathcal{D} = u^\mu \D_\mu$, $\sigma_{\mu\nu}=\mathcal{D}_\mu u_\nu+\mathcal{D}_\nu u_\mu$,  
and $\D_\mu$ denotes the Weyl-covariant
derivative, whose
action on the hydrodynamic fields is defined by their scaling under conformal
transformations; explicit formulae can be found in \cite{Loganayagam:2008is}.  The key
properties of the Weyl-covariant derivative are $\D_\mu u^\mu = 0$ and $u^\mu\D_\mu u_\nu = 0$.

The relaxation equations \eqref{eqs:relax_non} can now be written in the form\footnote{The conformal version of Eqs~\eqref{eqs:relax_non} is most easily obtained by explicitly separating the contribution to $\Pi$ that survives the conformal limit.}
\begin{subequations}
\be
&\tau_\A& 
\left[
\mathcal{D}\mathcal{A} 
+ \half\mathcal{A} \D\log\left(\f{\tau_\A}{\varepsilon \tau_\varphi T}\right)
\right]
+ \mathcal{A} 
= -\tau_\varphi \mathcal{D}\varepsilon , \\
&\tau_\Q& \left[
\D\mathcal{Q}^{\mu}
+ \half \Q^\mu \D\log\left(\f{\tau_\Q}{\varepsilon\tau_\psi T}\right)
\right]
+ \mathcal{Q}^\mu 
= -\tau_\psi \Delta^{\mu\lambda}\D_{\lambda}\varepsilon , 
\\
&\tau_\pi &
\left[
\mathcal{D}\pi^{\mu\nu} 
+ \half\pi^{\mu\nu} \D\log\left(\f{\tau_\pi}{\eta T}\right)
\right]
+ \pi^{\mu\nu} 
= - 2 \eta \sigma^{\mu\nu} ,
\ee
\label{eq:relax}
\end{subequations}
where $\tau_\A \equiv \varphi \beta_\A$, $\tau_\Q\equiv \psi \beta_\Q$, $\tau_\pi\equiv 2\eta \beta_\pi$ are three independent relaxation times,  and $\tau_\psi=\psi/(4\varepsilon)$, $\tau_\varphi=\varphi/(4\varepsilon)$ are two additional transport parameters. 
Conformal symmetry requires all relaxation times to scale as $1/T$.

A non-rotating global equilibrium state of a system described by 
the system of equations 
\eqref{eq:conservation0}, \eqref{eq:relax} 
is characterized by constant $\varepsilon$ and constant background velocity, i.e., $u^\mu = \gamma(1,\vv)$, $\vv$ is constant,  $\gamma = 1/\sqrt{1-|{\bf v}|^2}$, while $\mathcal{A}, \ \mathcal{Q}^\mu$, and $\pi^{\mu\nu}$ vanish. 
The behavior of small perturbations of such a state can be studied by considering small fluctuations $u^\mu \to u^\mu + \delta u^\mu$, $\varepsilon \to \varepsilon +\delta \varepsilon$ and similarly for the remaining fields $\mathcal{A}, \ \mathcal{Q}^\mu, \ \pi^{\mu\nu}$. 
One may represent fluctuations 
in Fourier space as $\delta \epsilon = \int d^4k/(2\pi)^4 e^{\Gamma t + ik_i  x^i} \delta \varepsilon (\Gamma, \kk)$, and similarly for the other variables. 

The linearized equations of motion determining the modes split into sound (longitudinal) and shear  (transverse) channels. In the local rest frame ($\mathbf{v}=0$), the dispersion relations of  shear modes
expanded in powers of $|\kk |$ up to $\mathcal{O}(|\kk|^2)$ read
\begin{align}
\Gamma^\perp_\text{h}=& - \frac{\eta}{sT} |\kk|^2, \\
\Gamma^\perp_{\text{nh},1}=& -\frac{1}{\tau_\pi} + \frac\eta{sT}\frac {(\tau_Q - \tau_\pi)}{(\tau_Q - \tau_\pi - 3\tau_\psi)} |\kk|^2 , \label{shearpi}\\
\Gamma^\perp_{\text{nh},2}=&-\frac{1}{\tau_Q - 3\tau_\psi} - \frac \eta {sT} \frac{3\tau_\psi}{(\tau_Q - \tau_\pi-3\tau_\psi)} |\kk|^2 .
\end{align}
Therefore,
in the shear channel there is a single hydrodynamic mode, $\Gamma^\perp_\text{h}$, and a pair of nonhydrodynamic (transient) modes, $\Gamma^\perp_{\text{nh},1}$ and $\Gamma^\perp_{\text{nh},2}$. In the sound channel, the dispersion relations expanded up to $\mathcal{O}(|\kk|^2)$ read
\begin{align}
\Gamma^\parallel_{\text{h},\pm} =& \pm i\frac{|\kk|}{\sqrt{3}}-\frac{\eta}{sT}\frac{2  |\kk|^2}{3
  }  , \\
\Gamma^\parallel_{\text{nh},1}=&-\frac1 {\tau_\pi}  +\frac{\eta}{sT} \frac{4(\tau_Q-\tau_\pi)}{3(\tau_Q-\tau_\pi-3\tau_\psi)}|\kk|^2, \label{soundpi}\\
\Gamma^\parallel_{\text{nh},2}=&-\frac{1}{\tau_A-\tau_{\varphi }}+\frac{  \tau_{\varphi } \tau_{\psi }}{ \tau_Q - \tau_A+\tau_{\varphi
   }-3\tau_{\psi }}|\kk|^2, \\
\Gamma^\parallel_{\text{nh},3}=&-\frac{1}{\tau_Q-3\tau _{\psi }} 
+\tau_\psi\frac{\frac{4\eta} {sT}(\tau_A-\tau_Q+3\tau_\psi - \tau_\varphi)+\tau_\varphi(\tau_\pi-\tau_Q+3\tau_\psi)}{(\tau_Q-\tau_\pi-3\tau_\psi)(\tau_Q-\tau_A-3\tau_\psi + \tau_\varphi)} |\kk|^2 . 
\end{align}
Thus, we have the usual two hydrodynamic sound modes, $\Gamma^\parallel_{\text{h},\pm}$, and three nonhydrodynamic modes, $\Gamma^\parallel_{\text{nh},1}$, $\Gamma^\parallel_{\text{nh},2}$, $\Gamma^\parallel_{\text{nh},3}$.  Note that in both channels we have additional nonhydrodynamic modes
as compared to the spectrum of linearized perturbations in  MIS~\cite{Baier:2007ix} as well as BDNK~\cite{Bemfica:2017wps} theories. In fact, when $\tau_\A=\tau_\Q=\tau_\psi=\tau_\varphi=0$ (MIS limit in the Landau frame) both the shear and the sound channel have only one nonhydrodynamic mode \cite{Baier:2007ix} and, in the BDNK limit, the shear channel has one nonhydrodynamic mode while the sound channel has two \cite{Bemfica:2017wps}.

Conditions to ensure linear causality  can be found by investigating the principal part of the linearized equations of motion~\cite{ChoquetBruhatGRBook}. We focus for simplicity on the case where $\tau_\A=\tau_\Q=\tau_\pi\equiv\ttt$. Following standard analyses employed in the calculation of characteristics of viscous hydrodynamic equations \cite{Bemfica:2017wps, Bemfica:2019knx,Bemfica:2020zjp}, ones finds the following necessary and sufficient
conditions for causality to hold in the linear regime for arbitrary background velocity:
$
0 \leq\eta/[sT(\ttt-3\tau_\psi)]\leq 1$,
 $b^2 - 4 c \geq 0$, and $0\leq \frac{1}{2}\left(b\pm\sqrt{b^2 - 4c}\right)\leq 1$,
where
\begin{eqnarray}
&&b=\frac{1}{3}\frac{\left[(\tau_R-\tau_\varphi)\left(\frac{4\eta}{sT}+\tau_R - 3\tau_\psi\right)+3\tau_\psi\tau_\varphi\right]}{(\tau_R-\tau_\varphi)(\tau_R-3\tau_\psi)},\\
&&c=\frac{1}{3}\tau_\psi\frac{\left(\frac{4\eta}{sT}+\tau_\varphi\right)}{(\tau_R-\tau_\varphi)(\tau_R-3\tau_\psi)}.
\end{eqnarray}
Additional details can be found in the Supplemental Material.
Using the dispersion relation analysis in the local rest frame, one can prove that stability holds (see Supplemental Material) if $\ttt>0$,  $\eta>0$,  $\ttt-3\tau_\psi >0$, $\ttt-\tau_\varphi >0$, $\tau_\psi>0$, $\tau_\varphi>0$.
For the range of parameters where the evolution is causal, one can prove that stability in the local rest frame implies stability in any other Lorentz frame \cite{Bemfica:2020zjp,Gavassino:2021owo}. One possible set of parameters which simultaneously satisfies the conditions for stability and causality found above is $T\tau_R = 5 \eta/s$ \cite{Denicol:2011fa,Denicol:2010xn} and $0<T\tau_\psi=T\tau_\varphi\leq \eta/s$. We note that the corresponding BDNK limit \cite{Bemfica:2017wps} is also causal and stable in this case. Therefore, one can see that both our second-order theory and its first-order truncation are causal and stable and, thus, amenable to simulations.

\mysection{Bjorken flow} A simple, but nontrivial class of highly-symmetric flows which is relevant to studies of the quark-gluon plasma created in heavy-ion collisions
was introduced by Bjorken~\cite{Bjorken:1982qr} and explored in numerous subsequent studies (see e.g. \cite{Florkowski:2017olj} for a review). In Bjorken flow all
hydrodynamic fields depend only on the proper time elapsed after the collision
and  $u^\mu=(1,0,0,0)$ in Milne coordinates where $x^\mu = (\tau,x,y,\varsigma)$,  $\tau=\sqrt{t^2-z^2}$, and $\varsigma = \tanh^{-1}(z/t)$. Keeping in mind applications to quark-gluon plasma, here we also consider the conformal case at zero chemical potential.

For Bjorken flow $\mathcal{Q}^\mu=0$ 
and the evolution equations reduce to a system of coupled first-order ordinary differential equations that determine $\varepsilon(\tau)$, its out-of-equilibrium correction $\A(\tau)$, and the relevant piece of the shear-stress tensor $\pi(\tau)$. 
We 
consider here the simplest case of our relaxation equations, where we drop the second term in brackets in each of \rf{eq:relax} and  
focus on the special case of equal relaxation times
$\tau_\A=\tau_\pi \equiv C_\tau/T$. We also set $\tau_\varphi=C_\varphi/T$ and $\eta=C_\eta s$, where $C_\tau, C_\varphi$, and $C_\eta$ are dimensionless constants as required by conformal symmetry. As in Refs.~\cite{Heller:2015dha,Aniceto:2015mto}, it is convenient to parameterize the phase space in terms of  dimensionless quantities; to this end we define 
\be
\xa = 6 \left(1 + \f{3}{4} \tau\p_\tau\ln{\varepsilon}\right),\quad
\xb = \f{\mathcal{A}}{\varepsilon}.
\ee
Regarded as functions of the dimensionless variable $w\equiv\tau T$, they satisfy
a pair of coupled differential equations
\be
\label{eq.A}
\frac{1}{12}(C_\tau - C_\varphi) w (\xa  + 12) \xa'
-\frac{3}{8} w \xa (\xb  - 4)
+\frac{(C_\tau - C_\varphi)}{3} \xa^2
-\frac{9}{2} w \xb - 12 C_\eta = 0
\ee
and
\be
\label{eq.B}
\f{1}{12} C_\tau w (\xa  + 12) \xb'
+ \f{1}{3}\xa ( C_\tau \xb+  C_\varphi) +
 \f{3}{2} w \xb  =0,
\ee
where the prime denotes differentiation with respect
to $w$. In the special case where $C_\varphi=0$ these equations admit a solution with
$\xb\equiv 0$ and then \eqref{eq.A} reduces to the equation satisfied by the
pressure anisotropy in MIS theory \cite{Florkowski:2017olj}. For $C_\varphi\neq 0$ it is however clear that
even if $\xb=0$ initially, it will be generated by the evolution. The late time 
asymptotics of solutions are $\xa \sim 8 C_\eta/w$ and $\xb \sim -63 C_\eta C_\varphi/27 w^2$ for all initial conditions. 
Computing higher order terms in these series one finds that they have a vanishing radius of convergence and form the basis of transseries solutions in a way similar to what happens in MIS theory \cite{Heller:2015dha,Basar:2015ava,Aniceto:2015mto}. 

\begin{figure}
\includegraphics[width=0.6\textwidth]{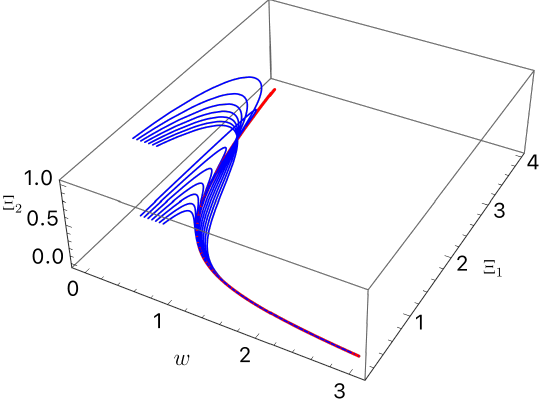}
\caption{The blue curves depict solutions whose initial
  conditions were set at several values of $w$ between $0.05$ and $0.3$. The red curve represents the attractor. The parameter values used when making the plot were $C_\eta = 0.08, C_\tau=0.2, C_\varphi=0.01$.}
  \label{fig:attr}
\end{figure}

Finally, we observe that Eqs.~\eqref{eq.A},~\eqref{eq.B} imply attractor behavior already at early times. This is reminiscent of the MIS attractor~\cite{Heller:2015dha}, but now in a $3$-dimensional phase space. 
Generic solutions decay to the attractor locus defined by the regular solution with initial conditions at $w=0$ given by 
\be
\xa(0) &=& 6 \sqrt{C_\eta/(C_\tau-C_\varphi)},\nn\\ \xb(0) &=& - C_\varphi/C_\tau.
\ee
This behavior can be seen in Fig.~\ref{fig:attr}. 

\mysection{Conclusions} We have formulated a new second-order theory of relativistic fluid dynamics which generalizes MIS theory by 
refraining from imposing any frame conditions. 
This introduces new transient nonhydrodynamic degrees of freedom, which could be eliminated if one truncates the gradient expansion, but are present in the full theory. On the conceptual side, our results uniquely clarify and unify the MIS and BDNK approaches by providing a consistent second-order framework which can be causal and stable even when truncated to first order in gradients. This is the first example of a theory of relativistic viscous hydrodynamics with this property. It would be interesting to work out how the second-order theory in a general frame proposed here can be derived from the Boltzmann equation using different approaches \cite{Denicol:2012cn,Rocha:2021lze,Tsumura:2015fxa}, or the  relaxation time approximation \cite{Rocha:2021zcw}.  

Conditions that ensure causality and stability for linearized perturbations in the conformal regime (at zero chemical potential) were presented in this work. Extending such linear analysis to the nonconformal case at nonzero chemical potential is conceptually straightforward. However, determining the conditions under which causality and hyperbolicity hold in the nonlinear far-from-equilibrium regime of the theory presented here will be considerably more challenging.


The new theory presented here can be readily applied in heavy-ion collision simulations as the new relaxation equations in \eqref{eq:relax} could be implemented in different numerical approaches (see, e.g. \cite{Schenke:2010nt,Karpenko:2013wva,Noronha-Hostler:2013gga,DelZanna:2013eua,Shen:2014vra,Romatschke:2017ejr,Pang:2018zzo}). In this case, one may still set initial conditions such that $u_\mu T^{\mu\nu} = -\varepsilon u^\nu$, but $\mathcal{A}$, $\mathcal{Q}^\mu$ and $\N$ will be nonzero throughout the subsequent evolution. It would be interesting to see the differences between our theory and BDNK in simulations of the quark-gluon plasma, especially at early times where deviations from equilibrium can be large.  
We also expect our results to be relevant for astrophysical applications, especially in the context of viscous neutron star merger simulations \cite{Alford:2017rxf,Shibata:2017jyf,Most:2021zvc}. Ongoing studies of the numerical properties of BDNK theories, such as the one recently performed in \cite{Pandya:2021ief,Pandya:2022pif,Bantilan:2022ech}, will also benefit from our approach, as it gives a clear prescription how to simulate first and second-order causal relativistic viscous hydrodynamics in a single unified framework. It would be interesting to extend the analysis done in \cite{Bantilan:2022ech} to include the theory proposed here.

\begin{acknowledgments}
\mysection{Acknowledgements} 
JN is partially supported by the U.S.~Department of Energy, Office of Science, Office for Nuclear Physics under Award No. DE-SC0021301.
MS is supported by the National Science Centre, Poland, under grants 2018/29/B/ST2/02457 and 2021/41/B/ST2/02909. For the purpose of Open Access, the author has applied a CC-BY public copyright licence to any Author Accepted Manuscript (AAM) version arising from this submission.

\end{acknowledgments}

\bibliography{References}{}
\bibliographystyle{bibstyl}

\newpage

\supplemental{SUPPLEMENTAL MATERIAL}

\subsection{Equations of motion in the general case}

The most general form for the entropy current reads
\begin{equation}
   \bar{S}_\text{neq}^\mu= S^\mu_\text{neq} + S^\mu_{\text{neq}, \perp}  ,
\end{equation}
where $S^\mu_\text{neq}$ is given by Eq. \eqref{eq:entrcur} and 
\begin{widetext}
\begin{equation}
    S^\mu_{\text{neq}, \perp} = \frac1T\left( \alpha_{\N\Q}\N\mathcal{Q}^\mu + \alpha_{\Pi \Q} \Pi\mathcal{Q}^\mu + \alpha_{\A\Q} \mathcal{A}\mathcal{Q}^\mu + \alpha_{\N \q} \N \q^\mu + \alpha_{\Pi \q} \Pi \q^\mu + \alpha_{\A \q} \mathcal{A}\q^\mu + \alpha_{\pi \Q}\pi^{\mu\nu}\mathcal{Q}_\nu + \alpha_{\pi \q}\pi^{\mu\nu} \q_\nu \right)
\end{equation}
\end{widetext}
contains off-equilibrium terms orthogonal to $u^\mu$ which were not considered in the main text of the paper. Using the equations of motion \eqref{eq:conservation}, the entropy production $\partial_\mu \bar{S}_{neq}^\mu$ is non-negative if the following relaxation equations for the dissipative quantities are satisfied
\begin{widetext}
\begin{subequations}
\begin{align}
 \frac{\mathcal{A}}{\varphi T} = &-\frac{\beta_\A}{T} D \mathcal{A} + D\left(\frac1T \right) -\frac12 \mathcal{A} \partial_\nu \left( \frac{\beta_\A}{T}u^\nu \right)  
  + \frac{\alpha_{\N\A}}{T} D \N + \gamma_{\N\A} \N D \left( \frac{\alpha_{\N\A}}{T}\right) + \tilde{\gamma}_{\N\A} \frac{\alpha_{\N\A}}{T} \N \theta \nonumber\\
 &+ \frac{\alpha_{\Pi \A}}{T} D \Pi + \gamma_{\Pi \A} \Pi D \left( \frac{\alpha_{\Pi \A}}{T}\right) + \tilde{\gamma}_{\Pi \A} \frac{\alpha_{\Pi \A}}{T} \Pi \theta + \frac{\alpha_{\A\Q}}{T} \partial_\nu \mathcal{Q}^\nu + \gamma_{\A\Q} \mathcal{Q}^\nu \partial_\nu \left( \frac{\alpha_{\A\Q}}{T} \right) \nonumber\\ 
 &+ \frac{\alpha_{\A\q}}{T} \partial_\nu \q^\nu + \gamma_{\A\q} \q^\nu \partial_\nu \left( \frac{\alpha_{\A\q}}{T} \right) ,
  \\
 \frac{\Pi}{\zeta T} = &-\frac{\beta_\Pi}{T} D \Pi -\frac 1T \theta -\frac12 \Pi \partial_\nu \left( \frac{\beta_\Pi}{T}u^\nu \right) + \frac{\alpha_{\N\Pi}}{T} D \N + \gamma_{\N\Pi} \N D \left( \frac{\alpha_{\N\Pi}}{T}\right) + \tilde{\gamma}_{\N\Pi} \frac{\alpha_{\N\Pi}}{T} \N \theta \nonumber\\
 &+ \frac{\alpha_{\Pi \A}}{T} D \mathcal{A} + (1- \gamma_{\Pi \A}) \mathcal{A} D \left( \frac{\alpha_{\Pi \A}}{T}\right) + (1-\tilde{\gamma}_{\Pi \A}) \frac{\alpha_{\Pi \A}}{T} \A \theta + \frac{\alpha_{\Pi \Q}}{T} \partial_\nu \mathcal{Q}^\nu + \gamma_{\Pi \Q} \mathcal{Q}^\nu \partial_\nu \left( \frac{\alpha_{\Pi \Q}}{T} \right) \nonumber\\ 
 &+ \frac{\alpha_{\Pi \q}}{T} \partial_\nu \q^\nu + \gamma_{\Pi \q} \q^\nu \partial_\nu \left( \frac{\alpha_{\Pi \q}}{T} \right) , \\
 \frac{\mathcal{Q}^\nu}{\psi T} = &-\frac{\beta_\Q}{T} \Delta^{\nu}_\lambda D \mathcal{Q}^\lambda + \Delta^{\nu}_\lambda\partial^{\lambda}\left(\frac1T\right) - \frac1T Du^\nu -\frac12 \mathcal{Q}^\nu \partial_\lambda \left( \frac{\beta_\Q}{T}u^\lambda \right) 
 + \frac{\alpha_{\Q \q}}{T} \Delta^{\nu}_\lambda D \q^\lambda + \gamma_{\Q\q} \q^\nu D \left( \frac{\alpha_{\Q \q}}{T} \right) 
  \nonumber\\
 & + \tilde{\gamma}_{\Q\q} \frac{\alpha_{\Q \q}}{T} \q^\nu \theta + \frac{\alpha_{\N \Q}}{T} \Delta^{\nu}_\lambda \partial^{\lambda} \N + \gamma_{\N\Q} \N \Delta^{\nu}_\lambda \partial^{\lambda} \left( \frac{\alpha_{\N \Q}}{T} \right)
  + \frac{\alpha_{\Pi \Q}}{T} \Delta^{\nu}_\lambda \partial^{\lambda} \Pi + (1-\gamma_{\Pi \Q}) \Pi \Delta^{\nu}_\lambda \partial^{\lambda} \left( \frac{\alpha_{\Pi \Q}}{T} \right) 
  \nonumber\\
  & + \frac{\alpha_{\A \Q}}{T}\Delta^{\nu}_\lambda \partial^{\lambda} \mathcal{A}  + (1-\gamma_{\A\Q}) \mathcal{A} \Delta^{\nu}_\lambda\partial^{\lambda} \left(\frac{\alpha_{\A \Q}}{T} \right) + \frac{\alpha_{\pi \Q}}{T} \Delta^{\nu}_\lambda\partial_\rho \pi^{\rho{\lambda}} + \gamma_{\pi \Q} \Delta^{\nu}_\lambda \pi^{\rho{ \lambda}}\partial_\rho \left( \frac{\alpha_{\pi \Q}}{T} \right) , \\
 \frac{\pi^{\nu\lambda}}{2\eta  T} = &-\frac{\beta_\pi}{T} \Delta^{\nu\lambda\delta\rho} D \pi_{\delta\rho} -\frac{\sigma^{\nu\lambda}}{T} -\frac12 \pi^{\nu\lambda} \partial_\rho \left( \frac{\beta_\pi}{T}u^\rho \right)  + \frac{\alpha_{\pi \Q}}{T}\Delta^{\nu\lambda\delta\rho} \partial_{\delta} \mathcal{Q}_{\rho} + (1-\gamma_{\pi \Q})\Delta^{\nu\lambda\delta\rho}\mathcal{Q}_{\delta} \partial_{\rho} \left(\frac{\alpha_{\pi \Q}}{T} \right)  \nonumber\\
 &  + \frac{\alpha_{\pi \q}}{T} \Delta^{\nu\lambda\delta\rho}\partial_{\delta} \q_{\rho} + \gamma_{\pi \q}\Delta^{\nu\lambda\delta\rho}\q_{\delta} \partial_{\rho} \left(\frac{\alpha_{\pi \q}}{T}\right) ,  \\
 \frac{\N}{\xi T} = & -\frac{\beta_\N}{T} D \N - D\left(\frac\mu T\right) -\frac12 \N \partial_\nu \left( \frac{\beta_\N}{T}u^\nu \right)  + \frac{\alpha_{\N\Pi}}{T} D \Pi + (1- \gamma_{\N\Pi}) \Pi D \left( \frac{\alpha_{\N\Pi}}{T} \right) + (1- \tilde{\gamma}_{\N\Pi}) \frac{\alpha_{\N\Pi}}{T} \Pi \theta 
  \nonumber\\
 & + \frac{\alpha_{\N\A}}{T} D \mathcal{A} + (1- \gamma_{\N\A}) \mathcal{A} D \left( \frac{\alpha_{\N\A}}{T} \right) + (1- \tilde{\gamma}_{\N\A}) \frac{\alpha_{\N\A}}{T} \mathcal{A} \theta + \frac{\alpha_{\N\Q}}{T} \partial_\nu \mathcal{Q}^\nu + (1-\gamma_{\N\Q})\mathcal{Q}^\nu \partial_\nu \left(\frac{\alpha_{\N\Q}}{T} \right)  \nonumber\\
 & +\frac{\alpha_{\N\q}}{T} \partial_\nu \q^\nu + (1-\gamma_{\N\q})\q^\nu \partial_\nu \left(\frac{\alpha_{\N\q}}{T} \right)  , \\
 \frac{\q^\nu}{\kappa T}= & -\frac{\beta_\q}{T} \Delta^\nu_\lambda D {\q}^\lambda - \Delta^\nu_\lambda \partial^{\lambda}\left(\frac \mu T\right)  -\frac12 \q^\mu \partial_\lambda \left( \frac{\beta_\q}{T}u^\lambda \right) + \frac{\alpha_{\Q \q}}{T}  \Delta^\nu_\lambda D\mathcal{Q}^\lambda + (1-\gamma_{\Q\q}) \mathcal{Q}^\nu D \left( \frac{\alpha_{\Q \q}}{T} \right) 
  \nonumber\\
 & + (1-\tilde{\gamma}_{\Q\q})\frac{\alpha_{\Q \q}}{T} \mathcal{Q}^\nu \theta  + \frac{\alpha_{\N \q}}{T} \Delta^\nu_\lambda \partial^{\lambda} \N + \gamma_{\N\q} \N \Delta^\nu_\lambda \partial^{\lambda} \left( \frac{\alpha_{\N \q}}{T} \right)
  + \frac{\alpha_{\Pi \q}}{T} \Delta^\nu_\lambda \partial^{\lambda} \Pi + (1-\gamma_{\Pi \q}) \Pi \Delta^\nu_\lambda \partial^{\lambda} \left(\frac{\alpha_{\Pi \q}}{T}\right) 
  \nonumber\\
  & + \frac{\alpha_{\A \q}}{T}\Delta^\nu_\lambda \partial^{\lambda}  \mathcal{A}  + (1-\gamma_{\A\q}) \mathcal{A} \Delta^\nu_\lambda \partial^{\lambda} \left(\frac{\alpha_{\A \q}}{T} \right) + \frac{\alpha_{\pi \q}}{T} \Delta^\nu_\lambda \partial_\rho \pi^{\rho{\lambda}} + (1-\gamma_{\pi Q}) \Delta^\nu_\lambda\pi^{\rho{\lambda}}\partial_\rho \left( \frac{\alpha_{\pi \q}}{T} \right) .
\end{align}
\label{eq:relful}
\end{subequations}
\end{widetext}
The $\gamma$ and $\tilde{\gamma}$-coefficients arise when redistributing terms in the entropy production such as, e.g., \cite{Hiscock_Lindblom_stability_1983}
\begin{equation}
    \N \A D \left( \frac{\alpha_{\N\A}}{T} \right) = \gamma_{\N\A} \N\A D \left( \frac{\alpha_{\N\A}}{T} \right) 
    + (1-\gamma_{\N\A})\N \A D \left( \frac{\alpha_{\N\A}}{T} \right).
\end{equation}
Using the conservation laws, one can show that the gradient expansion of Eqs.\ \eqref{eq:relful} truncated at first order is the same as Eqs.\ \eqref{eqs:gradexp}.

\subsection{Causality and linear stability in the conformal regime}
We study causality and stability in the case  $\tau_\A=\tau_\Q=\tau_\pi\equiv\ttt$. 
Causality \cite{ChoquetBruhatGRBook} places constraints on the transport coefficients. These constraints can be found by analyzing the system's characteristics obtained by solving the characteristic determinant associated with the 
principal part of the linearized equations of motion~\cite{ChoquetBruhatGRBook}. In the shear channel, this analysis leads to the following polynomial $(u_\mu \chi^\mu) [(u_\mu \chi^\mu)^2(\ttt-3\tau_\psi)-\eta/(sT)\Delta_{\mu\nu}\chi^\mu\chi^\nu]$, where $\chi_\mu$ is the covector normal to the characteristic hypersurface. Causality requires that the roots $\chi_\mu = (\chi_0(\chi_i),\chi_i)$ of the polynomial are real and $\chi_\mu \chi^\mu \geq 0$ \cite{Bemfica:2017wps,Bemfica:2020zjp}. For the shear channel one then finds
\begin{equation}
\label{causalityshear}
0 \leq\frac{\eta}{sT(\ttt-3\tau_\psi)}\leq 1.
\end{equation}
In the sound channel, the characteristic analysis gives the following polynomial  
\begin{equation}
(u_\alpha \chi^\alpha)\left[(u_\mu \chi^\mu)^2+c_1 \Delta_{\mu\nu}\chi^\mu\chi^\nu\right]\left[(u_\beta \chi^\beta)^2+c_2 \Delta_{\alpha\beta}\chi^\alpha\chi^\beta\right],
\end{equation} 
where 
$c_{1,2}=(1/2)(b\pm\sqrt{b^2-4c})$, with
\begin{eqnarray}
&&b=\frac{1}{3}\frac{\left[(\tau_R-\tau_\varphi)\left(\frac{4\eta}{sT}+\tau_R - 3\tau_\psi\right)+3\tau_\psi\tau_\varphi\right]}{(\tau_R-\tau_\varphi)(\tau_R-3\tau_\psi)},\\
&&c=\frac{1}{3}\tau_\psi\frac{\left(\frac{4\eta}{sT}+\tau_\varphi\right)}{(\tau_R-\tau_\varphi)(\tau_R-3\tau_\psi)}.
\end{eqnarray}
In order for causality to hold, $c_{1,2}$ have to be real and such that $0 \leq c_{1,2} \leq 1$, i.e.,
\begin{align}
\label{causalitysound1}
b^2 - 4 c &\geq 0, \\
\label{causalitysound2}
0\leq \frac{1}{2}\left(b\pm\sqrt{b^2 - 4c}\right)&\leq 1.
\end{align}

We now turn to the linear stability analysis in the local rest frame. Shear and sound disturbances are determined by the following equations
\begin{align}
\label{transversemode}
(\ttt \Gamma+1)\left[\Gamma^2(\ttt - 3\tau_\psi)+ \Gamma +\frac{\eta}{sT}|\kk|^2\right]&=0, \\
(\ttt \Gamma+1)(a_0\Gamma^4+a_1\Gamma^3+a_2\Gamma^2+a_3\Gamma+a_4)&=0, \label{longitudinalmode}
\end{align}
respectively, where the coefficients are:
    \begin{eqnarray}
a_0 &=& (\ttt-\tau_\varphi)(\ttt-3\tau_\psi),\\
a_1 &=& 2\ttt - \tau_\varphi-3\tau_\psi ,\\
a_2 &=& \frac{|\kk|^2}{3}\left[(\ttt-\tau_\varphi)\left(\frac{4\eta}{sT}+\ttt - 3\tau_\psi\right)+3\tau_\psi\tau_\varphi\right]+ 1 ,\\
a_3&=& \frac{|\kk|^{2}}{3}\left[\frac{4\eta}{sT}+2\ttt-\tau_\varphi-3\tau_\psi\right],\\
a_4&=& \frac{|\kk|^{2}}{3} + \frac{|\kk|^{4}}{3}\tau_\psi\left(\frac{4\eta}{sT}+\tau_\varphi\right).
\end{eqnarray}

Linear stability requires that
$\text{Re}(\Gamma)\leq 0$ for any constant and uniform background velocity $|\vv|$. The polynomials in the case $|\vv|\neq 0$ can be obtained by replacing  $\Gamma \to \gamma (\Gamma + i k_i v^i)$ and $|\kk|^2 \to -\gamma^2 (\Gamma + i k_i v^i)^2 + \Gamma^2 + |\kk|^2$.
In the shear channel \eqref{transversemode}, by following the same steps as in MIS theory \cite{Hiscock_Lindblom_stability_1983,Pu:2009fj,Brito:2020nou} and BDNK \cite{Bemfica:2017wps}, we conclude that stability holds if
\be
\ttt>0, \quad \eta>0, \quad \ttt-3\tau_\psi >0. 
\label{stabilityshear}
\ee 
For the sound channel we first consider the constraints coming from the $|\vv|=0$ and  $|\kk|=0$ case. It is easy to see that the constraints for stability are given by the third condition in 
\eqref{stabilityshear} and 
\be
\ttt-\tau_\varphi >0.
\label{stabilitysound1}
\ee
When $|\vv|=0$ and $|\kk|\neq0$, one can use the Routh-Hurwitz criterion \cite{gradshteyn2007} to find the conditions for stability of the fourth-order polynomial in \eqref{longitudinalmode}. These are given by \eqref{stabilityshear} and  \eqref{stabilitysound1} together with 
\begin{equation}
\label{stabilitysound22}
\tau_\psi>0, \quad \tau_\varphi>0.
\end{equation}
We note that the stability conditions \eqref{stabilityshear}, \eqref{stabilitysound1}, \eqref{stabilitysound22} imply $b>0$, $c>0$. Hence, the first inequality in \rf{causalitysound2} is automatically satisfied once stability and  \rf{causalitysound1} hold. Also, we note that the first inequality in \eqref{causalityshear} is satisfied if the stability conditions \eqref{stabilityshear} hold.

Let us now us discuss stability when $\mathbf{v}\neq 0$. It has been proved in Theorem III of Ref.\ \cite{Bemfica:2020zjp} that if the system is causal and strongly hyperbolic, stability in a given Lorentz frame implies stability in any other frame. More recently, it was shown in \cite{Gavassino:2021cli} that the assumption of strong hyperbolicity can be relaxed. This agrees with the general expectation that in a causal relativistic theory the stability of equilibrium is a Lorentz invariant concept. Therefore, the conditions stated above derived from causality and stability in the local rest frame are sufficient to guarantee stability for any Lorentz observer.   

As a consistency check, one may investigate the case when $\vv\neq0$ and $|\kk|= 0$, where the fourth-order polynomial in \eqref{longitudinalmode} factorizes as $\Gamma^2 P_2(\Gamma)$, where $P_2(\Gamma)$ is a quadratic polynomial. Using again the Routh-Hurwitz criterion, we find that the following conditions would imply stability in the boosted homogeneous case,
\begin{align}
\label{stabilitysound2}
2\ttt - \tau_\varphi-3\tau_\psi &> 2\frac{\eta}{sT},\\
\label{stabilitysound3}
(\ttt-\tau_\varphi)(\ttt-3\tau_\psi) &> \tau_\psi\tau_\varphi+2\frac{\eta}{sT}\left(\ttt-\tau_\varphi-\tau_\psi\right).
\end{align}
Indeed, one can numerically verify that the conditions above follow from the previous ones derived from causality and stability in the local rest frame.

\end{document}